\documentstyle[prd,aps,epsf]{revtex}
\newcommand{\segmentfig}[3]{
\begin{minipage}{#1}
\epsfxsize=#1
\epsffile{#2}
\begin{center}
{\small \mbox{#3}}
\end{center}
\end{minipage}}
\newcommand{\beq}{\begin{equation}}
\newcommand{\eeq}{\end{equation}}
\newcommand{\beqa}{\begin{eqnarray}}
\newcommand{\eeqa}{\end{eqnarray}}
\newcommand{\bea}{\begin{array}}
\newcommand{\ena}{\end{array}}
\begin{document}
\draft
\title{Have we already detected astrophysical symptoms of space-time 
noncommutativity ?}
\author{Takashi Tamaki
\thanks{electronic mail:tamaki@gravity.phys.waseda.ac.jp}, 
Tomohiro Harada,\thanks{electronic mail:harada@gravity.phys.waseda.ac.jp}
Umpei Miyamoto,\thanks{electronic mail:umpei@gravity.phys.waseda.ac.jp}
}
\address{Department of Physics, Waseda University,
Ohkubo, Shinjuku, Tokyo 169-8555, Japan} 
\author{and \\ 
Takashi Torii
\thanks{electronic mail:torii@resceu.s.u-tokyo.ac.jp}}
\address{Research Center for the Early Universe, 
University of Tokyo, Hongo, Bunkyo, Tokyo 113-0033, Japan
\\ and \\
Advanced Research Institute for Science and Engineering,
Waseda University, Ohkubo, 
Shinjuku, Tokyo 169-8555, Japan}
\date{\today}
\maketitle
\begin{abstract}
We discuss astrophysical implications of $\kappa$-Minkowski space-time, 
in which there appears space-time noncommutativity. 
We first derive a velocity formula for particles based on the motion of a 
wave packet. The result is that a massless particle moves 
at a constant speed as in the usual Minkowski space-time, which implies 
that an arrival time analysis by 
$\gamma$-rays from Markarian (Mk) 421 {\it does not} exclude 
space-time noncommutativity. 
Based on this observation, we analyze reaction processes in 
$\kappa$-Minkowski space-time which are related to the puzzling detections 
of extremely high-energy cosmic rays above the Greisen-Zatsepin-Kuzmin 
cutoff and of high-energy ($\sim$20 TeV) $\gamma$-rays from Mk 501. 
In these analyses, we take into account the ambiguity of the momentum 
conservation law which can not be determined uniquely 
from a mathematical viewpoint. We find that peculiar types of momentum 
conservation law with some length scale of noncommutativity above a 
critical length scale can explain such puzzling detections. 
We also obtain stringent constraints on the length scale of 
noncommutativity and the freedom of momentum conservation law. 
\end{abstract}

\section{Introduction and outline}

Recently, much attention has been paid to extremely high-energy cosmic rays 
(EHECRs) which have energies above that attained in any experimental 
apparatus on Earth\cite{Bird,Takeda}. It has been pointed out that 
these EHECRs provide an opportunity to investigate space-time 
properties on very short length scales or very high energy scales. 
The most striking feature is that some of these detections seem to be 
inconsistent with existing physics, in which such detections would be 
restricted by the Greisen-Zatsepin-Kuzmin (GZK) cutoff\cite{GZK}. 
That is, if we consider the interaction between EHECRs and CMB photons, 
particles with energy $\agt 7\times 10^{19}$ eV from distant sources 
cannot reach the Earth. There is also another anomalous phenomenon similar 
to this. That is detections of $\gamma$-rays above $\sim 20$ TeV from 
distant sources ($\agt 100$ Mpc) reported in Refs. \cite{Aharonian,Nikishov}. 
These $\gamma$-rays are expected to interact with infrared background 
(IRBG) photons and not to reach the Earth in a LI scenario\cite{Coppi}. 
In spite of exhaustive research, near sources which can explain such 
detections has not been found. 
Though there are many attempts explaining these anomalous phenomena, 
there is no absolute solution at present\cite{Olinto}. 

This could imply an encounter with new physics. Some authors argue 
that violation of Lorentz invariance (LI) might solve 
EHECRs above GZK cutoff\cite{Coleman,Jackiw,Bertolami,Aloisio,Sato}. 
LI violation might also explain detections of $\gamma$-rays above 
$\sim 20$ TeV. This possibility has been argued in Refs. 
\cite{Kifune,Kluzniak,Protheroe,Piran2}. 

One of the ways to introduce LI violation is to consider 
space-time noncommutativity with deformed LI, which has received 
attention in recent years since it naturally arises 
in the contexts of string/M theories
\cite{Douglas,Connes,Chu,Schomerus,Seiberg,Hashimoto}. It has also 
been argued that space-time uncertainty which comes from a 
fundamental string scale may be related to 
space-time noncommutativity\cite{Yoneya}. 

Apart from string/M theories, space-time noncommutativity also arises as a 
result of deformation 
quantization\cite{Jimbo}. Amelino-Camelia et al.\cite{Nature,Camelia,Piran} 
considered an interesting toy model called $\kappa$-Minkowski 
space-time where noncommutativity is introduced as 
$[x^{i},t]=i\lambda x^{i}$, 
where $\lambda$ is a free length scale and the index $i$ runs over 
$1,2,3$ \cite{Oeckl,MandO,Lukierski}. 
They obtained a severe constraint on $\lambda$ through an 
arrival time analysis of signals from a $\gamma$-ray 
burst\cite{Nature,Camelia}.  
If we accept this scenario, there is no room for detectable symptoms 
such as anomalous threshold to explain EHECRs\cite{Piran}. 

In these papers, the velocity of particles was evaluated using a 
group velocity formula in the usual Minkowski space-time. 
Here, we derive a more realistic velocity formula 
based on the motion of a wave packet 
in $\kappa$-Minkowski space-time. With this formula, 
we find that the space-time noncommutativity does not affect the 
velocity of massless particles. 
Motivated by this observation, we analyze reaction processes which are 
related to both detections of 
EHECRs beyond the GZK cutoff and of $\sim$20 TeV photons. 
In particular, we pay attention to the momentum conservation law which 
has some ambiguities in this model. We propose to determine the form 
of the momentum conservation law by deciding whether or not space-time 
noncommutativity is consistent with observations. In fact, we {\it can} 
exclude some forms of momentum conservation. 
Though our approach is purely kinematical, our result will provide 
a strong motivation to consider realistic model of space-time 
noncommutativity\cite{Carroll}. Throughout this paper, 
we use the units in which $c=\hbar =1$. 

\section{$\kappa$-Minkowski space-time}

We briefly review $\kappa$-Minkowski space-time. 
The basic commutation relations are 
\begin{eqnarray}
[x^{i},t]=i\lambda x^{i},\ \ \ [x^{i},x^{j}]=0, 
\label{eq:1}
\end{eqnarray}
where the indices $i,j$ run $1,2,3$. We can define differentiation, 
integration\cite{Oeckl} 
and Fourier transformation in this space-time\cite{MandO}. 
In order to define Fourier transformation consistently, 
the energy $E$ and the momentum $\mbox{\boldmath $p$}=(p_{1},p_{2},p_{3})$ 
of a particle form a non-Abelian group $G$ which can be written 
in a matrix form as, 
\begin{eqnarray}
(E,\mbox{\boldmath $p$}):=
\left(
\begin{array}{llll}
e^{\lambda E} & p_{1} & p_{2} & p_{3}  \\ 
0             &  1    &  0    &   0    \\  
0             &  0    &  1    &   0    \\  
0             &  0    &  0    &   1    
\end{array}
\right) \ .  
\label{eq:matrix}
\end{eqnarray}
Thus, if we denote the additive operator 
in $\kappa$-Minkowski space-time by $\hat{+}$ to distinguish 
from the conventional one, we can write as 
\begin{eqnarray}
(E_{1}\hat{+}E_{2},\mbox{\boldmath $p$}_{1}\hat{+}\mbox{\boldmath $p$}_{2})
&:=&(E_{1},\mbox{\boldmath $p$}_{1})(E_{2},\mbox{\boldmath $p$}_{2})
\nonumber  \\
&=&(E_{1}+E_{2},\mbox{\boldmath $p$}_{1}+e^{\lambda E_{1}}
\mbox{\boldmath $p$}_{2})\ \ .  \label{eq:3}
\end{eqnarray}
Following Ref.\cite{Camelia}, we describe a plane wave as 
\begin{eqnarray}
\psi_{(E,{\small\mbox{\boldmath $p$}})} =e^{i{\small\mbox{\boldmath $p$}\cdot 
\mbox{\boldmath $x$}}}e^{iEt},   \label{eq:6}
\end{eqnarray}
and place the $t$ generator to the right of $x$ generator, 
i.e., $\psi_{(E,{\small \mbox{\boldmath $p$}})} \neq 
e^{iEt}e^{i{\small \mbox{\boldmath $p$}\cdot \mbox{\boldmath $x$}}}$. 
Then, the property 
\begin{eqnarray}
\psi_{(E_{1},{\small \mbox{\boldmath $p$}_{1}})(E_{2},
{\small \mbox{\boldmath $p$}_{2}})}=\psi_{{\small \mbox{\boldmath $p$}_{1}},
E_{1}}\psi_{{\small \mbox{\boldmath $p$}_{2}},E_{2}}\ ,   \label{eq:wave}
\end{eqnarray}
is found. We can also define the wave in the reverse direction as 
\begin{eqnarray}
\psi_{(E,{\small\mbox{\boldmath $p$}})^{-1}}:=e^{-i
{\small\mbox{\boldmath $p$}}
e^{-\lambda E}\cdot {\small\mbox{\boldmath $x$}}}e^{-iEt}
=e^{-iEt}e^{-i{\small\mbox{\boldmath $p$}}\cdot 
{\small\mbox{\boldmath $x$}}} ,   
\label{eq:wave2}
\end{eqnarray}
which implies that $(E,\mbox{\boldmath $p$})^{-1}$ is an inversion 
of $(E,\mbox{\boldmath $p$})$. 

Because of these noncommutative structures, modification of 
Poincar\'e invariance is required 
to describe physics in a covariant way\cite{Lukierski}. 
The rotation and boost generators can be written as 
\begin{eqnarray}
M_{i}&=&-\epsilon_{imn}p_{m}\frac{\partial}{\partial p_{n}}, 
\label{eq:rotation}  \\
N_{i}&=&p_{i}\frac{\partial}{\partial E}-\left(
\frac{\lambda}{2}\mbox{\boldmath $p$}^{2}+\frac{1-e^{2\lambda E}}{2\lambda}
\right)\frac{\partial}{\partial p_{i}}+\lambda p_{i}p_{j}
\frac{\partial}{\partial p_{j}} .
\label{eq:boost}
\end{eqnarray}
Using (\ref{eq:boost}), a finite boost transformation for the $i=1$ 
direction can be obtained as\cite{Bruno}
\begin{eqnarray}
p_{1}&=&\frac{\tanh (\lambda m)\sinh \xi}{\lambda [1-\tanh (\lambda m)
\cosh\xi]}, \label{eq:Bruno1}  \\
p_{2}&=&p_{3}=0, \label{eq:Bruno2}  \\  
E&=&m+\frac{1}{\lambda}\ln \left[
\frac{1-\tanh (\lambda m)}{1-\tanh (\lambda m)\cosh \xi}
\right] ,\label{eq:Bruno3}
\end{eqnarray}
where $\xi$ is a boost parameter and we choose $\mbox{\boldmath $p$}=0$ 
and $E=m$, i.e., $m$ is a rest mass of the particle, for $\xi =0$. 

Because of (\ref{eq:boost}), the dispersion relation is altered as 
\begin{eqnarray}
\lambda^{-2}(e^{\lambda E}+e^{-\lambda E}-2)-\mbox{\boldmath $p$}^{2}
e^{-\lambda E}=K^{2}, 
\label{eq:5}
\end{eqnarray}
where $K$ is a constant with the dimension of mass. 
If we take a rest frame, this can be expressed as 
\begin{eqnarray}
\lambda^{-2}(e^{\lambda m}+e^{-\lambda m}-2)=K^{2}.  
\label{eq:5d}
\end{eqnarray}

\section{The velocity formula}

Here, we derive a new velocity formula which is one of the main  
results of this paper. The velocity of the particle in the 
usual Minkowski space-time is 
\begin{eqnarray}
\mbox{\boldmath $v$}=\frac{dE}{d\mbox{\boldmath $p$}} . 
\label{eq:7}
\end{eqnarray}
If we apply this in $\kappa$-Minkowski space-time, 
$|\mbox{\boldmath $v$}|=e^{-\lambda E}$ is obtained for a massless particle, 
where we used Eq.~(\ref{eq:5}). This formula, together with the data 
on $\gamma$-rays associated with Markarian (Mk) 421 in 
Ref.~\cite{Biller} leads to the constraint $|\lambda |\alt 10^{-33}$ meter 
\cite{Nature,Camelia,Biller}. 
Since this discussion depends crucially on the form of 
Eq.~(\ref{eq:7}), i.e., on what is the velocity, 
we reexamine the group velocity formula by 
forming a wave packet in $\kappa$-Minkowski space-time as a more realistic 
situation. For this purpose, we consider infinitesimal changes $\Delta E$ 
and $\Delta \mbox{\boldmath $p$}$ in $E$ and $\mbox{\boldmath $p$}$, 
respectively, as a result of adding 
($\Delta E',\Delta \mbox{\boldmath $p$}'$) as
\begin{eqnarray}
(E\hat{+}\Delta E',\mbox{\boldmath $p$}\hat{+}\Delta\mbox{\boldmath $p$}')
=(E+\Delta E,\mbox{\boldmath $p$}+\Delta \mbox{\boldmath $p$}).  
\label{eq:9}
\end{eqnarray}
In this case, we can express $(\Delta E',\Delta \mbox{\boldmath $p$}')$ as 
\begin{eqnarray}
(\Delta E',\Delta \mbox{\boldmath $p$}')\cong (\Delta E,\frac{\Delta 
\mbox{\boldmath $p$}}{e^{\lambda E}}),  
\label{eq:9d}
\end{eqnarray}
where we keep only terms in first order in $\Delta E$ and 
$\Delta \mbox{\boldmath $p$}$. By using Eqs.~(\ref{eq:wave}) and 
(\ref{eq:9d}), we make a wave packet as follows\cite{footnote2} 
\begin{eqnarray}
I&=&\psi_{(E-\Delta E,{\small\mbox{\boldmath $p$}}-\Delta 
{\small\mbox{\boldmath $p$}})}+\psi_{(E+\Delta E,{\small\mbox{\boldmath $p$}}
+\Delta {\small\mbox{\boldmath $p$}})}   
\nonumber \\
&=&\psi_{(E,{\small\mbox{\boldmath $p$}})(-\Delta E',-\Delta 
{\small\mbox{\boldmath $p$}}')}+\psi_{(E,{\small\mbox{\boldmath $p$}})
(\Delta E',\Delta {\small\mbox{\boldmath $p$}}')} 
\nonumber  \\
&=&\psi_{(E,{\small\mbox{\boldmath $p$}})}\psi_{(-\Delta E',-\Delta 
{\small\mbox{\boldmath $p$}}')}+\psi_{(E,{\small\mbox{\boldmath $p$}})}
\psi_{(\Delta E',\Delta {\small\mbox{\boldmath $p$}}')}
\nonumber  \\
&=&\psi_{(E,{\small\mbox{\boldmath $p$}})}[e^{-i\Delta
{\small\mbox{\boldmath $p$}}'\cdot{\small\mbox{\boldmath $x$}}}
e^{-i\Delta E't}+e^{i\Delta{\small\mbox{\boldmath $p$}}'\cdot
{\small\mbox{\boldmath $x$}}}e^{i\Delta E't}]  \nonumber  \\
&\cong &2e^{i{\small\mbox{\boldmath $p$}}\cdot{\small\mbox{\boldmath $x$}}}
e^{iEt}\cos \left[\frac{\Delta \mbox{\boldmath $p$}}{e^{\lambda E}}\cdot
\left(\mbox{\boldmath $x$}+\frac{e^{\lambda E}\Delta Et}{\Delta 
\mbox{\boldmath $p$}}\right)\right] .\label{eq:11}
\end{eqnarray}
By considering $|I|^{2}$, the group velocity $\mbox{\boldmath $v$}_{l}$ 
can be written as 
\begin{eqnarray}
\mbox{\boldmath $v$}_{l}:=e^{\lambda E}\frac{dE}{d\mbox{\boldmath $p$}} .
\label{eq:12}
\end{eqnarray}

We also consider a similar relation 
\begin{eqnarray}
(\Delta E'\hat{+}E,\Delta\mbox{\boldmath $p$}'\hat{+}\mbox{\boldmath $p$})
=(E+\Delta E,\mbox{\boldmath $p$}+\Delta \mbox{\boldmath $p$}),  
\label{eq:9c}
\end{eqnarray}
which is different from Eq.~(\ref{eq:9}) due to noncommutativity. 
In this case, the corresponding group velocity 
$\mbox{\boldmath $v$}_{r}$ is
\begin{eqnarray}
\mbox{\boldmath $v$}_{r}:=\frac{\frac{dE}{d{\small\mbox{\boldmath $p$}}}}{
1-\lambda {\small\mbox{\boldmath $p$}}\cdot\frac{dE}{
d{\small\mbox{\boldmath $p$}}}
} .\label{eq:12c}
\end{eqnarray}
Using (\ref{eq:5}) and (\ref{eq:5d}), we obtain the important conclusion 
that {\it massless particles 
move in a constant speed $|\mbox{\boldmath $v$}_{l}|=
|\mbox{\boldmath $v$}_{r}|=1$ as in the usual Minkowski space-time 
for arbitrary $\lambda$}\cite{footnote1}. 
Therefore, the argument in Ref.\cite{Nature,Camelia,Biller}
does not apply. In this case, there appears the possibility that 
the large value of $\lambda$ ($\agt 10^{-33}$m) may 
solve the puzzling problems of EHECRs above 
GZK cutoff and of $\sim$20 TeV photons simultaneously. We investigate this 
possibility next. However, we emphasis on the importance of the result 
{\it not} because $\kappa$-Minkowski space-time can avoid the constraint 
{\it but} because our result provides an opportunity to reconsider 
LI deformation models in general. 

\section{Threshold anomaly}

We first consider the two-body head-on collision of particles and subsequent 
creation of two particles $1+2\to 3+4$. 
We define the energy $E_{i}$ and momentum $\mbox{\boldmath $p$}_{i}$ 
of the $i$-th particle as those in the laboratory frame. We denote 
the rest mass of the $i$-th particle as $m_{i}$. We also assume that $m_{2}=0$, 
$m_{3}\neq 0$, $m_{4}\neq 0$ 
and $\mbox{\boldmath $p$}_{i}=(p_{i},0,0)$. In the usual Minkowski 
space-time, we use the dispersion relation 
\begin{eqnarray}
E_{i}^{2}-p_{i}^{2}=m_{i}^{2}, 
\label{eq:t1}
\end{eqnarray}
and the energy momentum conservation law, 
\begin{eqnarray}
p_{1}+p_{2}&=&p_{3}+p_{4}, 
\label{eq:t2}   \\
E_{1}+E_{2}&=&E_{3}+E_{4}, 
\label{eq:t3}
\end{eqnarray}
to obtain the threshold value of $E_{1}$, which we denote by $E_{th,0}$. 
We assume that the resultant particles are at rest 
in the center-of-mass frame 
in the threshold reaction. In the laboratory frame, this means that the 
resultant particles move in the same speed, that is 
\begin{eqnarray}
\frac{p_{3}}{m_{3}}=\frac{p_{4}}{m_{4}} .
\label{eq:t4}
\end{eqnarray}
We also assume that $p_{2}$ has an opposite sign against that of $p_{1}$. 
If we neglect higher order terms in $E_{2}$, then 
\begin{eqnarray}
E_{th,0}=\frac{(m_{3}+m_{4})^{2}-m_{1}^{2}}{4E_{2}}. 
\label{eq:t5}
\end{eqnarray}
We also define the threshold value of $p_{1}$ as $p_{th,0}$ which 
can be approximated as $p_{th,0}\sim E_{th,0}$. 

Next, we consider the same reaction in $\kappa$-Minkowski space-time. 
Eq.(\ref{eq:t1}) is replaced by 
\begin{eqnarray}
&&\lambda^{-2}(e^{\lambda E_{i}}+e^{-\lambda E_{i}}-2)
-(p_{i})^{2}e^{-\lambda E_{i}} \nonumber  \\ 
&&=\lambda^{-2}
(e^{\lambda m_{i}}
+e^{-\lambda m_{i}}-2) .  
\label{eq:t6}
\end{eqnarray}
If we interpret the algebra in $\kappa$-Minkowski space-time faithfully, 
the energy momentum conservation law is expressed as 
\begin{eqnarray}
(E_{1},p_{1})(E_{2},p_{2})=(E_{3},p_{3})(E_{4},p_{4}) .  
\label{eq:2}
\end{eqnarray}
Even if it holds, one should note that we need a 
rule to distinguish two particles. If we 
consider the collision of two particles with $A$, $B\in G$, respectively, 
does it correspond to $AB$, $BA$ or anything else ? 
At present, we have no way to determine it. Amelino-Camelia 
et al.\cite{Camelia} proposed to find the rule by experiments. 
Here, we introduce a phenomenological parameter $a$, which 
controls the form of conservation law as follows: 
\begin{eqnarray}
&&a(E_{1},p_{1})(E_{2},p_{2})+(1-a)(E_{2},p_{2})(E_{1},p_{1}) 
\nonumber \\
&=&a(E_{3},p_{3})(E_{4},p_{4})+(1-a)(E_{4},p_{4})(E_{3},p_{3})\ . 
\label{eq:2c}
\end{eqnarray}
As regards plausible values for $a$, care must be taken. 
If we consider two particles of the same species, $a=1/2 $ would 
be physically reasonable value, since if they have same energy and move 
opposite direction each other, they have zero total momentum 
only for this choice. In fact, the parameter $a$  
may be a function of physical quantities of two particles 
such as mass, charge and/or spin for two different species. 
Here, we use the same value of $a$ on the left and the 
right hand sides of (\ref{eq:2c}) for convenience. 
Moreover, we restrict our attention to $0\leq a\leq 1$ for clarity. 

We also need to impose the condition that the resultant particles are at 
rest in the center-of-mass frame. To obtain a relation between momenta 
$p_{3}$ and $p_{4}$, we use the boost transformation (\ref{eq:Bruno1}). 
For the same value of $\xi$, we obtain 
\begin{eqnarray}
\frac{p_{3}}{\tanh (\lambda m_{3})}=\frac{p_{4}}{\tanh (\lambda m_{4})}\ . 
\label{eq:ratio}
\end{eqnarray}
We can solve $E_{1}$ as a function of $a$, $\lambda$, $m_{1}$, $m_{3}$, 
$m_{4}$ and $E_{2}$ by using (\ref{eq:t6}), (\ref{eq:2c}) and 
(\ref{eq:ratio}). We apply this result to two astrophysical cases. 

\subsection{Threshold anomaly for TeV $\gamma$-rays}

Here, we consider the process $\gamma +\gamma \to e^{+}+e^{-}$, 
which may occur when a $\gamma$-ray travels in the IRBG. 
In this case, $m_{1}=0$ and $m_{3}=m_{4}=m_{e}$, where 
$m_{e}$ is the electron mass. If we assume the existence of IRBG 
photons ($0.2\alt E_{2} \alt 5$ eV) then the threshold is 
$E_{th,0}\sim 1$ TeV in Minkowski space-time. Then, the reported detection 
of $\sim$20 TeV photons from Mk501 ($\sim$150 Mpc from the Earth) 
would be difficult to explain\cite{Aharonian}. 

We summarize the equation for the threshold in $\kappa$-Minkowski 
space-time which is derived from (\ref{eq:t6}), (\ref{eq:2c}) and 
(\ref{eq:ratio}) as 
\begin{eqnarray}
AB=yx(yx+1)^{2}\sinh^{2}\frac{\lambda m_{e}}{2},
\label{basic}
\end{eqnarray}
where 
\begin{eqnarray}
A&:=& (1-a)y^{4}-(1-2a)y^{2}-a ,\label{eq:g2d} \\
B&:=& ax^{4}+(1-2a)x^{2}+a-1 ,\label{eq:g2e}
\end{eqnarray}
and $x:=e^{\lambda E_{1}/2}$ and $y:=e^{\lambda E_{2}/2}$.  
Since we are considering the collision of two particles of the same species, 
$a=1/2$ would be physically reasonable. 
Note that, though we have $E_{th,0}\approx p_{th,0}$ for 
high energy particles in the usual Minkowski space-time, this is not 
the case in $\kappa$-Minkowski space-time. 

We should recall that, to estimate the energy of primary particles, 
we calculate the sum of energy of all secondary particles. 
Since energy is conserved in the usual sense even in $\kappa$-Minkowski 
space-time, we do not need to take into account the effect of space-time 
noncommutativity to estimate the energy of primary particles. 
Thus, the observation of $\sim$20 TeV photons in usual Minkowski 
space-time has the same meaning also in $\kappa$-Minkowski space-time. 
On the other hand, the usual sum 
of momenta of all secondary particles does not coincide with the momentum 
of the primary particle in this 
space-time. Therefore, if $p_{th}$ could be evaluated independently of the 
observation of $E_{th}$, it might become important to extract information 
about space-time noncommutativity through the detection of violation 
of the usual momentum conservation. We exhibit properties of both the 
energy and the momentum from this reason. 

We first show the dependence of $E_{th}$ and $p_{th}$ on $\lambda >0$ 
in Figs.~\ref{gamma1} (a) and (b), respectively. For simplicity, $E_{2}$ 
is chosen as $E_{2} =1$ eV for IRBG photons. For $a=0$, $E_{th}$ and $p_{th}$ 
increase with $\lambda$, compared with the same quantities in Minkowski 
space-time. In particular, $E_{th}$ and $p_{th}$ diverge for 
$\lambda :=\lambda_{c}\sim 4$ TeV$ ^{-1}$. That is, the universe is 
entirely tranparent for $\lambda >\lambda_{c}$. For $a=1/2$ and $1$, 
$p_{th}$ increases with $\lambda$, though $E_{th}$ decreases. 

For all $a$, a first-order correction in $\lambda$ arises for $p_{th}$. 
If we expand $p_{th}$ as $p_{th}=\sum_{k=0}^{\infty}\frac{p_{th,k}}
{k!}\lambda^{k}$, the first-order coefficient $p_{th,1}$ is written as 
\begin{eqnarray}
p_{th,1}=p_{th,0}\left[p_{th,0}(1-a)+E_{2}\left(a-\frac{1}{2}
\right)\right] . 
\label{eq:first}
\end{eqnarray}
On the other hand, the first-order correction in $\lambda$ for $E_{th}$, 
which we denote $E_{th,1}$, is written as 
\begin{eqnarray}
E_{th,1}=E_{th,0}(E_{th,0}-E_{2})\left(\frac{1}{2}-a\right) . 
\label{eq:firstE}
\end{eqnarray}
Thus, it disappears for $a=1/2$. 

The reason why $E_{th}$ and $p_{th}$ 
disappear for $a=0$ above $\lambda_{c}\sim 4$ TeV$ ^{-1}$ can be understood 
as follows. For $\lambda E_{th}\gg 1$, and $\lambda m_{e}$, 
$\lambda E_{2}$ $\ll 1$, we can approximate eq. (\ref{basic}) as 
\begin{eqnarray}
(1+\lambda E_{2})x&\approx &\lambda E_{2}-\frac{\lambda^{2}m_{e}^{2}}{2} 
\ \ \ {\rm for}\ \ a=0, \label{eq:approxa0}  \\  
aE_{2}x&\approx &\frac{\lambda m_{e}^{2}}{4} 
\ \ \ {\rm for}\ \ a\neq 0. \label{eq:approxan0}
\end{eqnarray}
In this range of approximation, since $\lambda E_{th,0}=\lambda m_{e}^{2}
/E_{2}$ is expected to be larger than $1$, eq.(\ref{eq:approxa0}) has no 
real solution $E_{th}$ while a real solution $E_{th}$ exists for $a\neq 0$. 
This means that $\lambda_{c}$ for $a=0$ is characterized by 
$1/E_{th,0}\sim 1$ TeV$ ^{-1}$. 

For $\lambda E_{th}$, $\lambda m_{e}$, $\lambda E_{2}$ $\gg 1$, 
we can summarize the results as follows. For $a=0$, eq. (\ref{basic}) is 
approximated as $y^{4}x^{2}\sim (yx)^{3}e^{\lambda m_{e}}/4$, which yields 
$E_{1}=-2m_{e}+E_{2}<0$. This contradicts the first assumption. So a 
solution does not exist. In a similar way, we can show that $E_{1}$ 
approaches $2m_{e}+E_{2}$ and $2m_{e}-E_{2}$ for $a=1$ and for $a\neq 0,1$, 
respectively. 

To investigate properties for $\lambda <0$, we replace $\lambda$ with 
$-\lambda$. In eq. (\ref{basic}), this corresponds to the replacement 
$x\to 1/x$ and $y\to 1/y$. We find that eq. (\ref{basic}) becomes 
invariant if $a$ is also replaced by $(1-a)$. 

Thus, the case $a\ll 1$, $\lambda\agt 4$ TeV$ ^{-1}$, 
and the case $1-a \ll 1$, $-\lambda\agt 4$ TeV$ ^{-1}$ remain as candidate 
solutions for $\sim$20 TeV photons. On the other hand, we can 
exclude $a=O(1)$ and $\lambda \agt 10$ TeV$ ^{-1}$, or $(1-a)=O(1)$ and 
$-\lambda \agt 10$ TeV$ ^{-1}$ from the present experimental data.

\subsection{Threshold anomaly for GZK cutoff}

Here, we consider the interaction of ultra high energy protons 
with CMB photons ($\sim 10^{-3}$eV) which results in a pair production 
$p+\gamma \to p+\pi_{0}$. In this case, $m_{1}=m_{3}=m_{p}$ and 
$m_{4}=m_{\pi}$, where $m_{p}$ and $m_{\pi}$ are the proton mass and 
the pion mass, respectively. Because of $E_{th,0}\sim 
7\times 10^{19}$ eV, it is difficult for EHECRs above $E_{th,0}$ to 
reach the Earth from cosmologically distant sources. 

\begin{figure}
\begin{center}
\segmentfig{12cm}{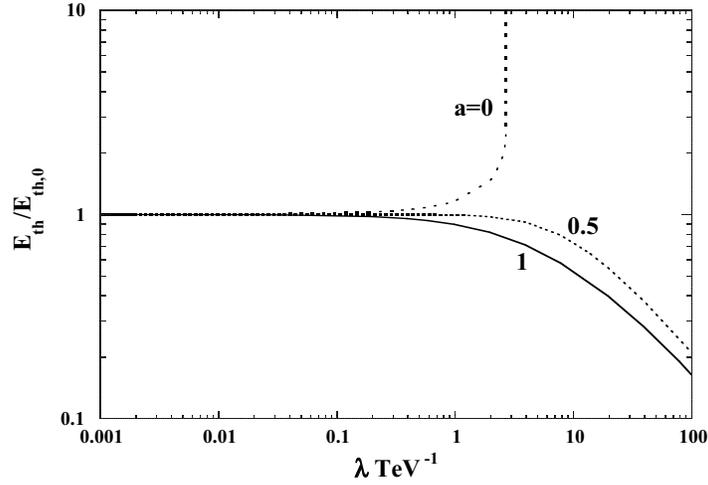}{(a)}  \\
\segmentfig{12cm}{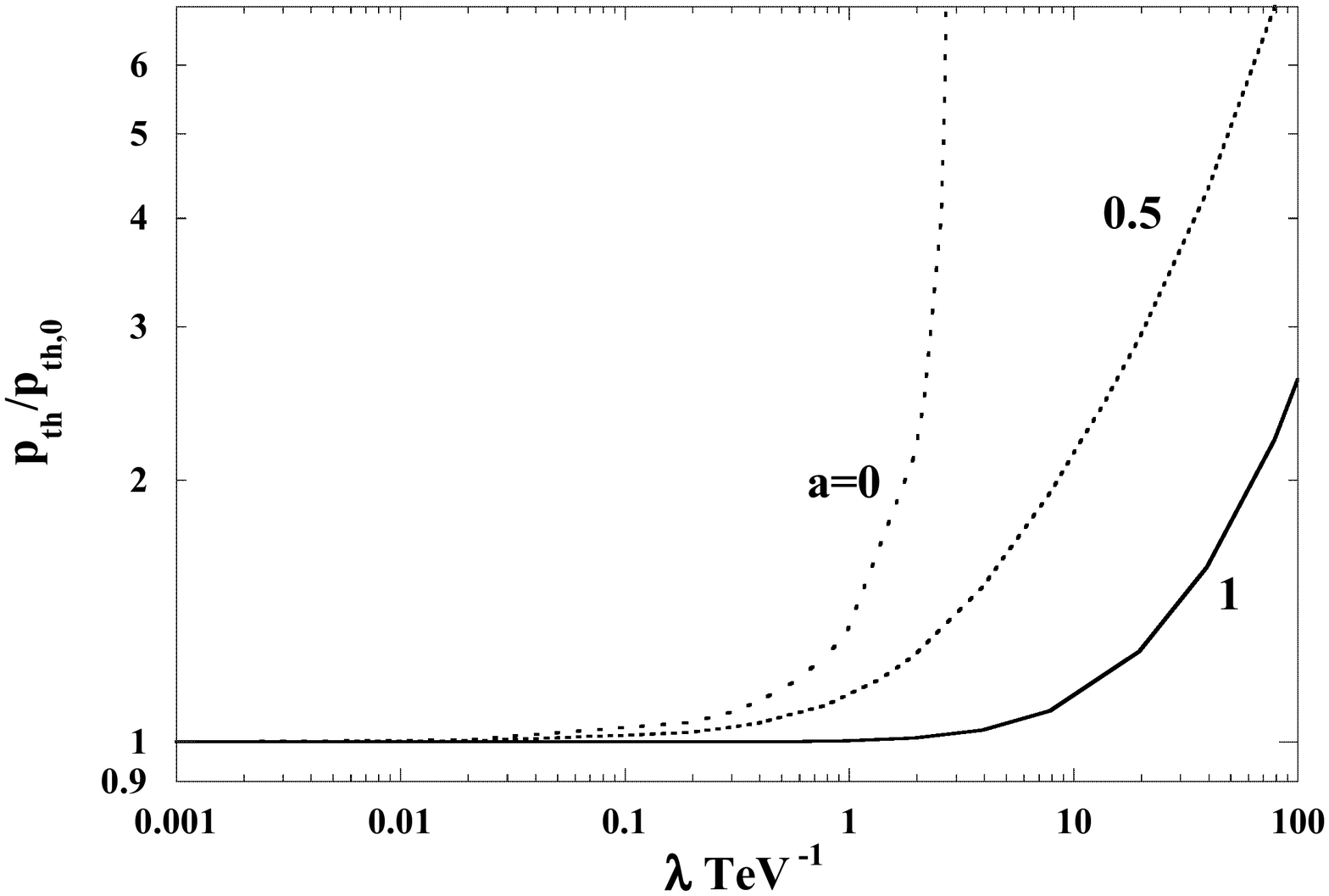}{(b)}
\caption{Threshold anomaly for TeV-$\gamma$ rays for $\lambda >0$. 
(a) $\lambda$-$E_{th}$, (b) $\lambda$-$p_{th}$ 
are plotted for $E_{2} =1$ eV. For $a=0.5$ and $1$, 
$E_{th}$ decreases with $\lambda$ increases for $\lambda >1$ TeV$ ^{-1}$, 
while $p_{th}$ monotonically increases. The
$a\ll 0$ case is only desirable to explain $\sim$20 TeV photons. 
It is noted that $E_{th}$ is invariant under the transformation 
$\lambda\to -\lambda$ and $a\to (1-a)$. 
\label{gamma1} }
\end{center}
\end{figure}

\begin{figure}
\begin{center}
\segmentfig{14cm}{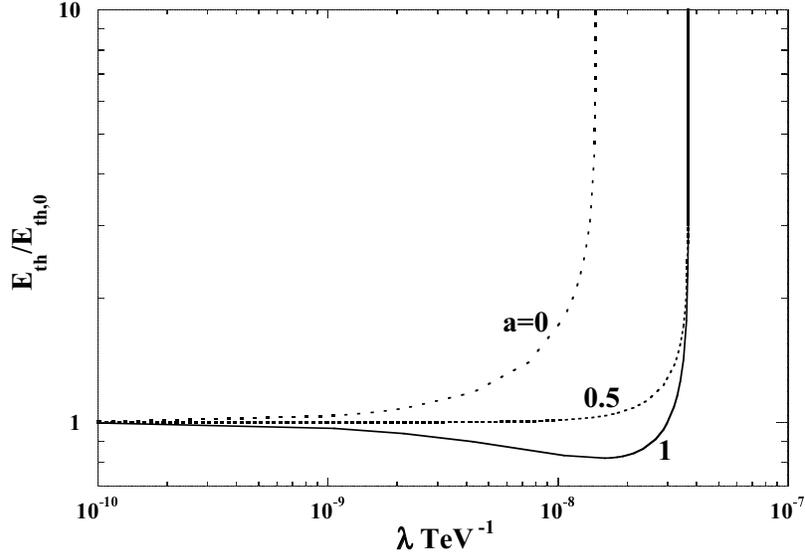}{(a)}  \\
\segmentfig{14cm}{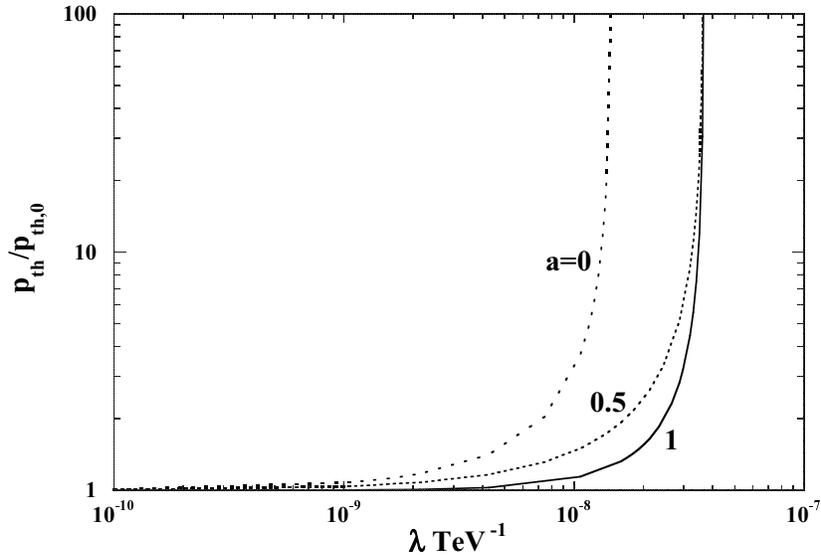}{(b)}
\caption{Threshold anomaly for GZK cutoff for $\lambda >0$. 
(a) $\lambda$-$E_{th}$, (b) $\lambda$-$p_{th}$ are 
plotted for $E_{2} =10^{-3}$ eV. Though qualitative 
features for small $\lambda$ are similar to those of Fig.~1, they show 
drastic difference from Fig.~1 for $\lambda \protect\agt 3\times 10^{-8}$ 
TeV$ ^{-1}$. 
\label{GZKfinal} }
\end{center}
\end{figure}

We solve (\ref{eq:t6}), (\ref{eq:2c}) and (\ref{eq:ratio}) numerically. 
We show the dependence of $E_{th}$ and $p_{th}$ for $\lambda >0$, 
in Figs.~\ref{GZKfinal} (a) and (b), respectively. 
$E_{2}$ is chosen as $E_{2} =10^{-3}$ eV for CMB photons. 
Compared with Fig.~\ref{gamma1}, we find that the qualitative features 
for small $\lambda$ are quite similar, i.e., $E_{th}/E_{th,0}>1$ for 
$a=0$ and $E_{th}/E_{th,0}<1$ for $a=1$, while $p_{th}/p_{th,0}>1$ for 
all cases. 

However, we find qualitative differences from Fig.~\ref{gamma1} for 
$\lambda \agt 10^{-8}$ TeV$ ^{-1}$. 
The threshold disappears for $\lambda > \lambda_{c}\sim 2\times 10^{-8}$ 
TeV$ ^{-1}$ in the $a=0$ case, which can be explained 
as in the $\gamma$-ray case since $\lambda_{c}$ coincides approximately 
with $1/E_{th,0}$. For the case $a=1/2$ and $1$, 
$E_{th}/E_{th,0}$ increases with $\lambda (\agt 3\times 10^{-8}$ 
TeV$ ^{-1}$) and disappears for $\lambda \agt 5\times 10^{-8}$ 
TeV$ ^{-1}$, unlike the $\gamma$-ray case. 

\begin{figure}
\begin{center}
\segmentfig{14cm}{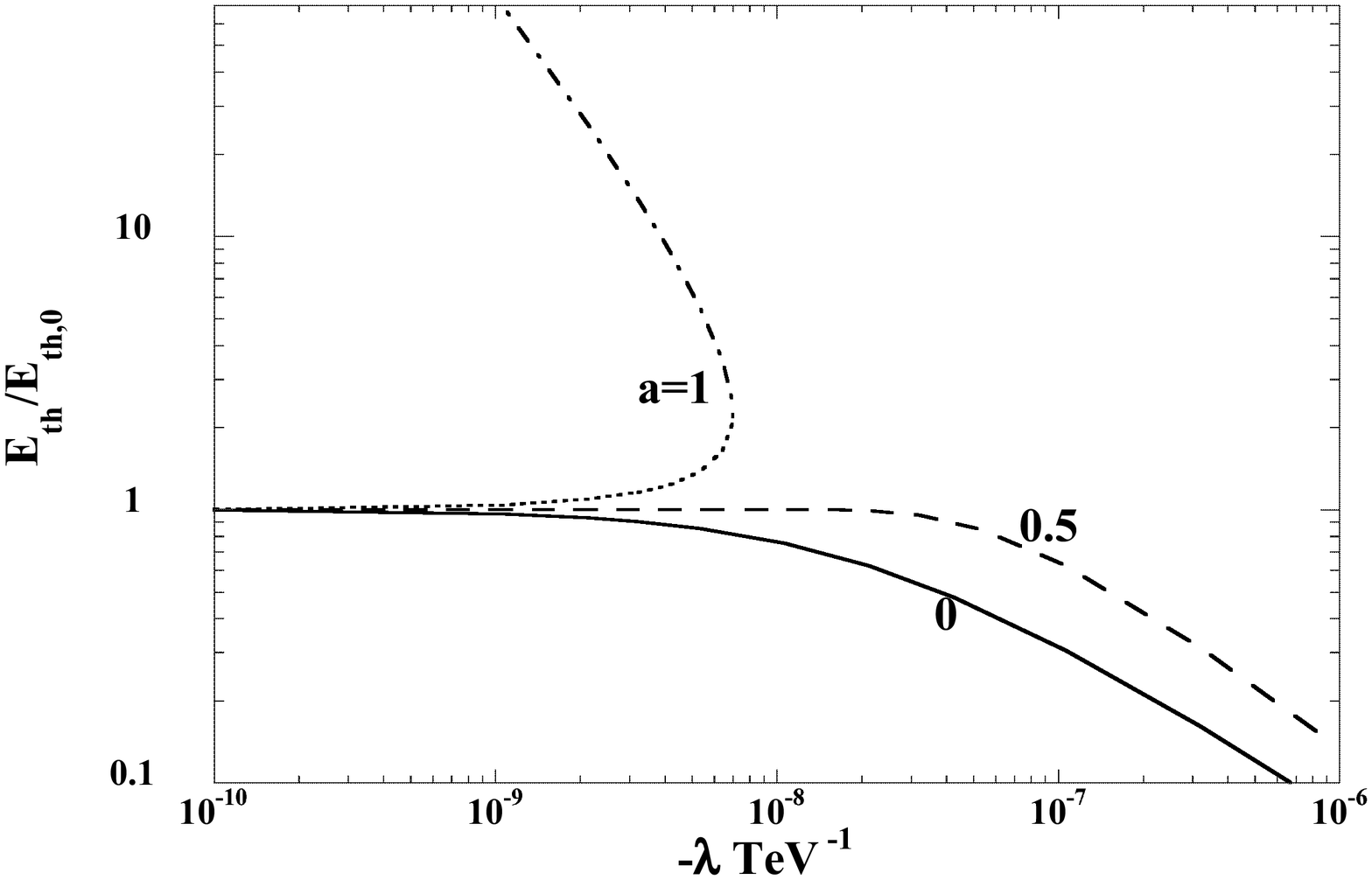}{(a)}  \\
\segmentfig{14cm}{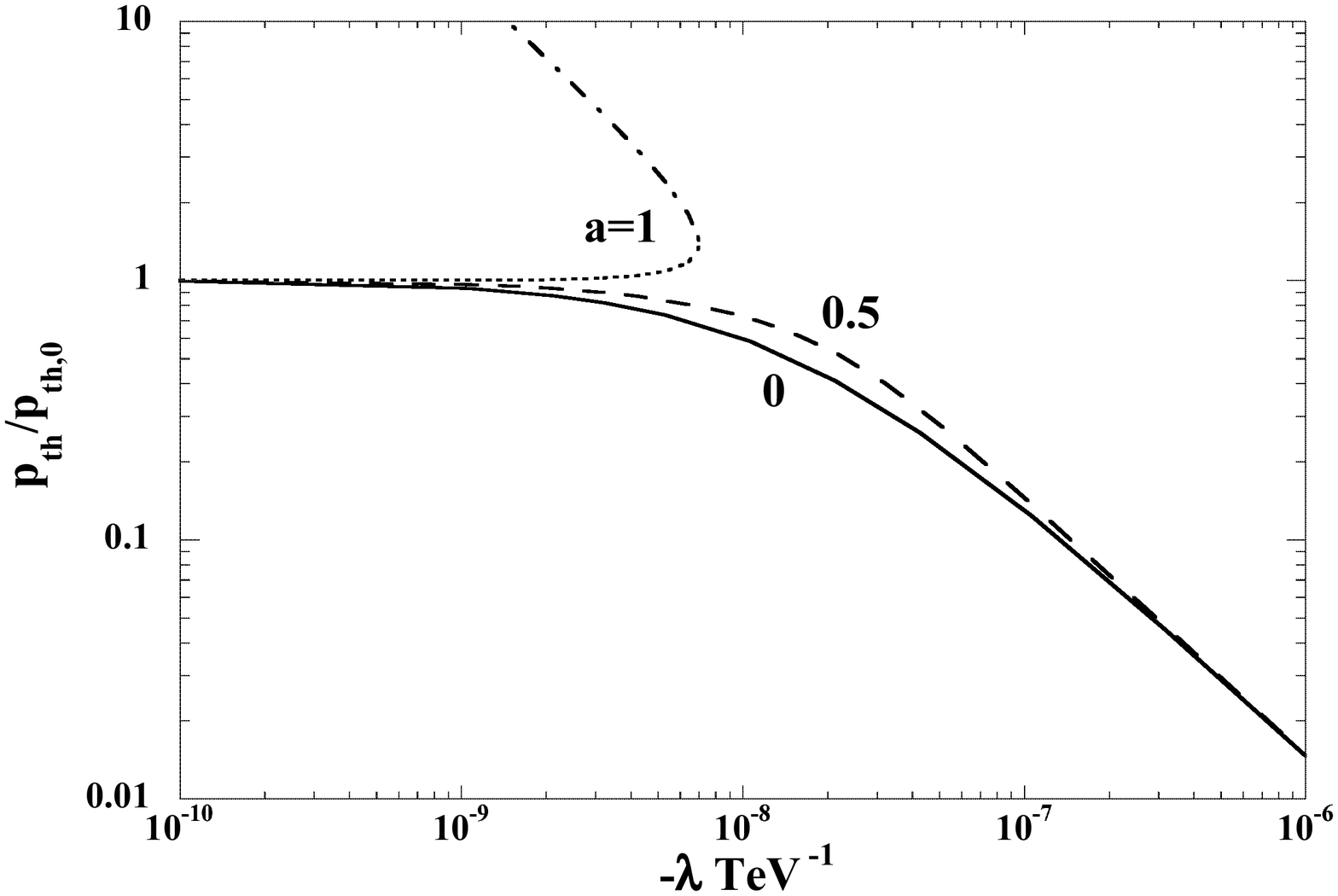}{(b)}
\caption{Threshold anomaly for GZK cutoff for $\lambda <0$. 
(a) $\lambda$-$E_{th}$, (b) $\lambda$-$p_{th}$ are 
plotted for $E_{2} =10^{-3}$ eV and $\lambda <0$. Unlike the case 
$\lambda >0$, the threshold vanishes only the $a=1$ case for 
$\lambda \protect\alt -7\times 10^{-9}$ TeV$ ^{-1}$.
\label{GZK2} }
\end{center}
\end{figure}

In this case, there is no simple symmetry about $\lambda \to -\lambda$, 
as found in the previous case. Thus, we also show the dependence 
of $E_{th}$ and $p_{th}$ for $\lambda <0$, in Figs.~\ref{GZK2} 
(a) and (b), respectively. We have a crucial difference from the 
$\gamma$-ray case even for $\lambda >-10^{-9}$ TeV$ ^{-1}$. For $a=1$, 
there is a value $E_{th2}$ over which the reaction does 
{\it not} occur. We denote $E_{th2}$ by a dot-dashed line. $E_{th2}$ 
diverges as $\lambda \to -0$ and merges with $E_{th}$ at 
$\lambda =\lambda_{c}\sim -7\times 10^{-9}$ TeV$ ^{-1}$. 

We find that the behavior for small $|\lambda |$ is 
that $E_{th}/E_{th,0}>1$ for $a=1$ and 
$E_{th}/E_{th,0}<1$ for $a=0$ as in the $\gamma$-ray case 
for $\lambda <0$. For $a=1/2$ and $1$, $E_{th}$ decreases 
with $\lambda (\alt -5\times 10^{-8}$ TeV$ ^{-1}$). 

Thus, the $a\ll 1$ case for $\lambda \agt 2\times 10^{-8}$ TeV$ ^{-1}$, 
the $a= O(1)$ case for $\lambda \agt 5\times 10^{-8}$ TeV$ ^{-1}$ and 
the $(1-a)\ll 1$ case for $\lambda\alt -7\times 10^{-9}$ TeV$ ^{-1}$
remains as candidate explanations for detections of super GZK events. 
For $(1-a)=O(1)$, we can exclude $\lambda \alt -10^{-7}$ TeV$ ^{-1}$. 

\section{Conclusion and discussion}

We have first considered a velocity formula to describe the particle 
motion based on the motion of a wave packet in $\kappa$-Minkowski space-time. 
In this formula, space-time noncommutativity does not affect 
the motion of a massless particle. Thus, an arrival time analysis 
of $\gamma$-ray bursts in Refs. \cite{Nature,Camelia,Biller} does 
not exclude space-time noncommutativity in this model. 
Since this feature had not been discussed so far, it should be 
stressed and is one of our main conclusions here. 

Based on this consideration, we have obtained 
threshold values for reactions $\gamma+\gamma \to e^{+}+e^{-}$ and 
$p+\gamma \to p+\pi_{0}$ in $\kappa$-Minkowski space-time and analyzed 
their relavance to the puzzling observations of $\sim$20 TeV photons 
and EHECRs above the GZK 
cutoff, introducing a parameter $a$ to take into account the ambiguity 
of the momentum conservation law. 

In the TeV $\gamma$-ray case, though $a=1/2$ is 
favorable in the physical context, only $a\ll 1$ for 
$\lambda\agt 4$ TeV$ ^{-1}$, or $(1-a)\ll 1$ for $\lambda\alt -4$ 
TeV$ ^{-1}$ appear able to explain the detections of $\gamma$-rays above 
$\sim$20 TeV. The possibilities $a= O(1)$ for $\lambda\agt 10$ TeV$ ^{-1}$, 
or $(1-a)=O(1)$ for $\lambda\alt -10$ TeV$ ^{-1}$, are excluded. 

In the EHECR case, we cannot assign definite values to $a$, because it may 
depend on, e.g., masses and/or charges of two particles. The possibilities 
$a\ll 1$ and $\lambda \agt 2\times 10^{-8}$ TeV$ ^{-1}$, or 
$a=O(1)$ and $\lambda \agt 5\times 10^{-8}$ TeV$ ^{-1}$ 
remain viable. We can exclude cases in which $(1-a)=O(1)$ and 
$\lambda \alt -10^{-7}$ TeV$ ^{-1}$. 

Thus, $a\ll 1$ for $\lambda\agt 4$ TeV$ ^{-1}$ or $(1-a)\ll 1$ 
for $\lambda\alt -4$ TeV$ ^{-1}$ appear able to explain both phenomena. 
Our results are important because they suggest that extremely high-energy 
particles might be expected in realistic models with space-time 
noncommutativity. If this is the case, then we might have already 
detected symptoms of the space-time noncommutativity. 

\section*{Acknowledgments}

Special thanks to Kei-ichi Maeda for continuous encouragement and to James 
Overduin for checking our English.  
This work was supported partly by the Grant-in-Aid (No.05540) from 
the Japanese Ministry of Education, Culture, Sports, Science and Technology, 
and partly by the Waseda University Grant for Special Research Projects.


\begin{thebibliography}{99}
\bibitem{Bird}
D. J. Bird et al., Astrophys. J. {\bf 441}, 144 (1995). 
\bibitem{Takeda}
M. Takeda et al., Phys. Rev. Lett. {\bf 81}, 1163 (1998). 
\bibitem{GZK}
K. Greisen, Phys. Rev. Lett. {\bf 16}, 748 (1966); G. T. Zatsepin 
and V. A. Kuzmin, Sov. Phys. JETP Lett. {\bf 4}, 78 (1966).  
\bibitem{Aharonian}
F. A. Aharonian et al., Astron. Astrophys. {\bf 349}, 11A (1999). 
\bibitem{Nikishov}
A. N. Nikishov, Sov. Phys. JETP {\bf 14}, 393 (1962); 
J. Gould, G. Schreder, Phys. Rev. D {\bf 155}, 1404 (1967); 
F. W. Stecker, O. C. De Jager and M. H. Salmon, Astro. Phys. J. 
{\bf 390}, L49 (1992). 
\bibitem{Coppi}
P. S. Coppi and F. A. Aharonian, Astropart. Phys. {\bf 11}, 35 (1999). 
\bibitem{Olinto}
See, e.g., A. V. Olinto, Phys. Rep. {\bf 333-334},329 (2000). 
\bibitem{Coleman}
S. Coleman and S. L. Glashow, Phys. Rev. D {\bf 59}, 116008 (1999). 
\bibitem{Jackiw}
R. Jackiw and V. A. Kostelecky, Phys. Rev. Lett. {\bf 82}, 3572 (1999). 
\bibitem{Bertolami}
O. Bertolami and C. S. Calvalho, Phys. Rev. D {\bf 61}, 103002 (2000). 
\bibitem{Aloisio}
R. Aloisio, P. Blasi, P. L. Ghia and A. F. Grillo, 
Phys. Rev. D {\bf 62}, 053010 (2000). 
\bibitem{Sato}
H. Sato and T. Tati, Prog. Theor. Phys. {\bf 47}, 1788 (1972); 
H. Sato, {\it Proc. of the International Workshop: Space Factory on 
JEM/ISS} (1999), astro-ph/0005218. 
\bibitem{Kifune}
T. Kifune, Astrophys. J. Lett. {\bf 518}, L21 (1999). 
\bibitem{Kluzniak}
W. Kluzniak, {\it Proc. of the 8th Internatinal Workshop on 
"Neutrino Telescopes"}, (Milla Baldo Ceolin ed., 1999), astro-ph/9905308. 
\bibitem{Protheroe}
R. J. Protheroe and H. Meyer, Phys. Lett. B {\bf 493}, 1 (2000). 
\bibitem{Piran2}
G. Amelino-Camelia and T. Piran, Phys. Rev. D {\bf 64}, 036005 (2001). 
\bibitem{Douglas}
M. R. Douglas and C. Hull, JHEP {\bf 02}, 8 (1998). 
\bibitem{Connes}
A. Connes, M. R. Douglas and A. Schwarz, JHEP {\bf 02}, 3 (1998). 
\bibitem{Chu}
C. S. Chu and P. M. Ho, Nucl. Phys. B {\bf 550}, 151 (1999). 
\bibitem{Schomerus}
V. Schomerus, JHEP {\bf 06}, 030 (1999). 
\bibitem{Seiberg}
N. Seiberg and E. Witten, {\bf 09}, 032 (1999). 
\bibitem{Hashimoto}
A. Hashimoto and N. Itzhaki, Phys. Lett. B {\bf 465}, 142 (1999). 
\bibitem{Yoneya}
T. Yoneya, {\it Wandering in the Fields}, p.419 (World Scientific, 1987); 
{\it ibid}., {\it Quantum String Theory}, p.23 (Springer, 1988); 
{\it ibid}., Prog. Theor. Phys. {\bf 103}, 1081 (2000). 
\bibitem{Jimbo}
M. Jimbo, Lett. Math. Phys. {\bf 10}, 63 (1985). 
\bibitem{Nature}
G. Amelino-Camelia et al., Nature {\bf 393}, 763 (1998); 
L. J. Garay, Phys. Rev. Lett. {\bf 80}, 2508 (1998); 
R. Gambini and G. Pullin, Phys. Rev. D {\bf 59}, 124021 (1999). 
\bibitem{Camelia}
G. Amelino-Camelia, J. Lukierski and A. Nowicki, 
Int. J. Mod. Phys. A {\bf 14}, 4575 (1999); 
G. Amelino-Camelia and S. Majid, Int. J. Mod. Phys. A {\bf 15}, 
4301 (2000). 
\bibitem{Piran}
G. Amelino-Camelia and T. Piran, Phys. Lett. B {\bf 497}, 265 (2001); 
G. Amelino-Camelia, J. Lukierski and A. Nowicki, hep-th/0103227. 
\bibitem{Carroll}
S. M. Carroll et al., Phys. Rev. Lett. {\bf 87}, 141601 (2001). 
\bibitem{Oeckl}
R. Oeckl, J. Math. Phys. {\bf 40}, 3588 (1999). 
\bibitem{MandO}
S. Majid and R. Oeckl, Commun. Math. Phys. {\bf 205}, 617 (1999). 
\bibitem{Lukierski}
J. Lukierski, A. Nowicki, H. Ruegg and V. N. Tolstoy, 
Phys. Lett. B {\bf 264}, 331 (1991); 
S. Majid and H. Ruegg, Phys. Lett. B {\bf 334}, 348 (1994). 
\bibitem{Bruno}
N. R. Bruno, G. Amelino-Camelia and J. Kowalski-Glikman, hep-th/0107039. 
\bibitem{Biller}
Biller et al., Phys. Rev. Lett. {\bf 83}, 2108 (1999). 
\bibitem{footnote2}
Strictly speaking, this is not a Gaussian wave packet. However, it is 
sufficient to obtain a group velocity. 
\bibitem{footnote1}
Recently, a particle velocity was also discussed in Ref.
\protect\cite{Glikman} which gives the same answer as 
ours for massless particles. 
\bibitem{Glikman}
J. Kowalski-Glikman, hep-th/0107054. 
\end{thebibliography}
\end{document}